\documentclass[runningheads]{svmult}

\usepackage{makeidx}   
\usepackage{graphicx}  
\usepackage{subeqnar}  
\usepackage{multicol}  
\usepackage{physprbb}  
\makeindex             

\begin{document}
\title*{The Stellar Population of High Redshift Galaxies}
\toctitle{The Stellar Population of High Redshift Galaxies}
%
%
\titlerunning{The Stellar Population of High Redshift Galaxies}
%
\author{D\"orte Mehlert\inst{1}
\and Stefan Noll\inst{1}
\and Immo Appenzeller \inst{1}
\and the FDF team\inst{1,2,3,4}}
\authorrunning{D\"orte Mehlert et al.}
%
%

\institute{$^1$ Landessternwarte Heidelberg, 
$^2$ Universit\"ats-Sternwarte M\"unchen,\\
$^3$       Universit\"ats-Sternwarte G\"ottingen,
$^4$       ESO Garching,
$^5$       MPIA Heidelberg}

\maketitle              


\vspace{-0.5cm}
\begin{multicols}{2}

\section{The Project}
\vspace{-0.25cm}
\noindent
The consortium of the institutes listed above, which combined forces to 
built the FORS instruments at the ESO VLT used a significant 
fraction of their guaranteed observing time to observe a ``FORS Deep Field''. 
One of the scientific objectives of the FDF is 
to study the dependence of the physical properties of galaxies
on the cosmic age. Hence we obtained deep multi-band images of the 
FDF as well as spectra of a subsample of galaxies in the FDF.
With these spectroscopic data we will investigate the 
stellar population (age of the starburst, IMF, metallicity, dust reddening)
of distant galaxies with the aim of deriving new information on the 
evolution of the young universe. 

\vspace{-0.4cm}
\section{The Observations}
\vspace{-0.25cm}
\noindent
During 3 nights of MOS observations with FORS2 at the VLT we obtained 
389 object spectra. 
Using the grism 150I and a slitwidth of 1'' we covered the spectral range from
$\lambda = 3000 ... 9200$~\AA\ with a spectral scale of 5 \AA /pixel. 
Depending on the object magnitude the integration times ranged between 2 and 
10 hours with average
seeing of 0.69''. Standard reduction (bias subtraction, 
flatfielding, cosmic ray elimination, sky subtraction, 
rebinning, etc.), was performed using MIDAS routines.

\vspace{-0.4cm}
\section{First Results}
\vspace{-0.25cm}
\noindent
For $\approx$ 203 objects we have spectra with sufficient S/N to determine the 
type and redshift. Among these we found 169 galaxies, 71 with $z > 1$.
Fig.~1 shows a typical spectrum of a high redshift galaxy ($z = 2.437$) 
in our sample. The most prominent feature is the Ly$_{\alpha}$ absorption line.
Additionally several metal absorption lines can be identified. Furthermore 
the intense (rest frame) UV continua indicates intensive starburst activity.
\vbox{
\noindent
\begin{center}
{\includegraphics[width=5.8cm]{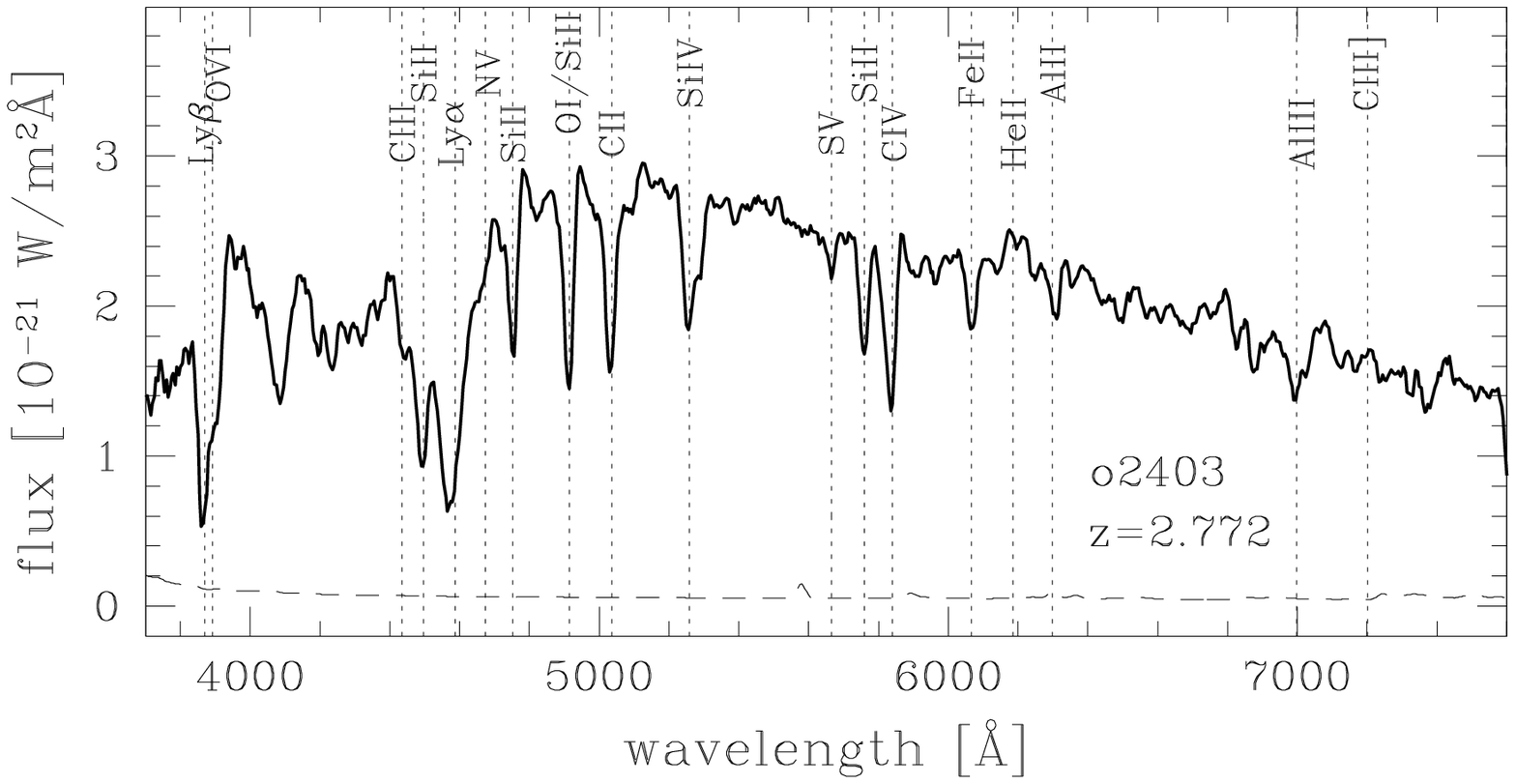}}
\end{center}
\vspace{-0.12cm}
{{\bf\scriptsize Fig. 1:}
\footnotesize
\em
Spectrum of a distant ($z=2.772$)
galaxy in the FDF with a S/N $\approx 35$ per resolution element (= 5 pixel). 
The dashed 
line indicates the noise level, which varies with wavelength due to the 
night sky spectrum and the wavelength dependent instrumental efficiency.
The position of some prominent spectral lines are indicated by the vertical 
dotted lines.
}}
\noindent
In Fig.~2 we present a comparison of three of our galaxies with synthetic 
spectra from Leitherer et al. (2001), showing the spectral region of the 
CIV\ resonance line. For galaxy 1747 ($z = 2.37$) and 2403 ($z = 2.77$) 
the spectral features of CIV are 
well represented by the solar and LMC model, respectively. On the other hand,
for~galaxy~2160 ($z = 3.26$) neither~the~solar~nor~the~LMC models fit the 
observed feature. 
Hence we conclude that the metallicity 
of object 2160 is much lower than the LMC value.\\
\noindent
Note that with increasing redshift the galaxies' metallicity seems to 
decrease. 
To investigate this trend for the whole sample we measured the equivalent 
width ($EW$) of the two prominent stellar absorption lines CIV and SiIV.
To increase the sample we included measurements for some galaxies in the 
field of the lensing cluster 1E0657, the HDF-S and the AXAF Deep Field, which  
we had 
observed during FORS commissioning runs with the same spectroscopic setup 
as described above.
\vbox{
\noindent
\begin{center}
{\includegraphics[width=5.8cm]{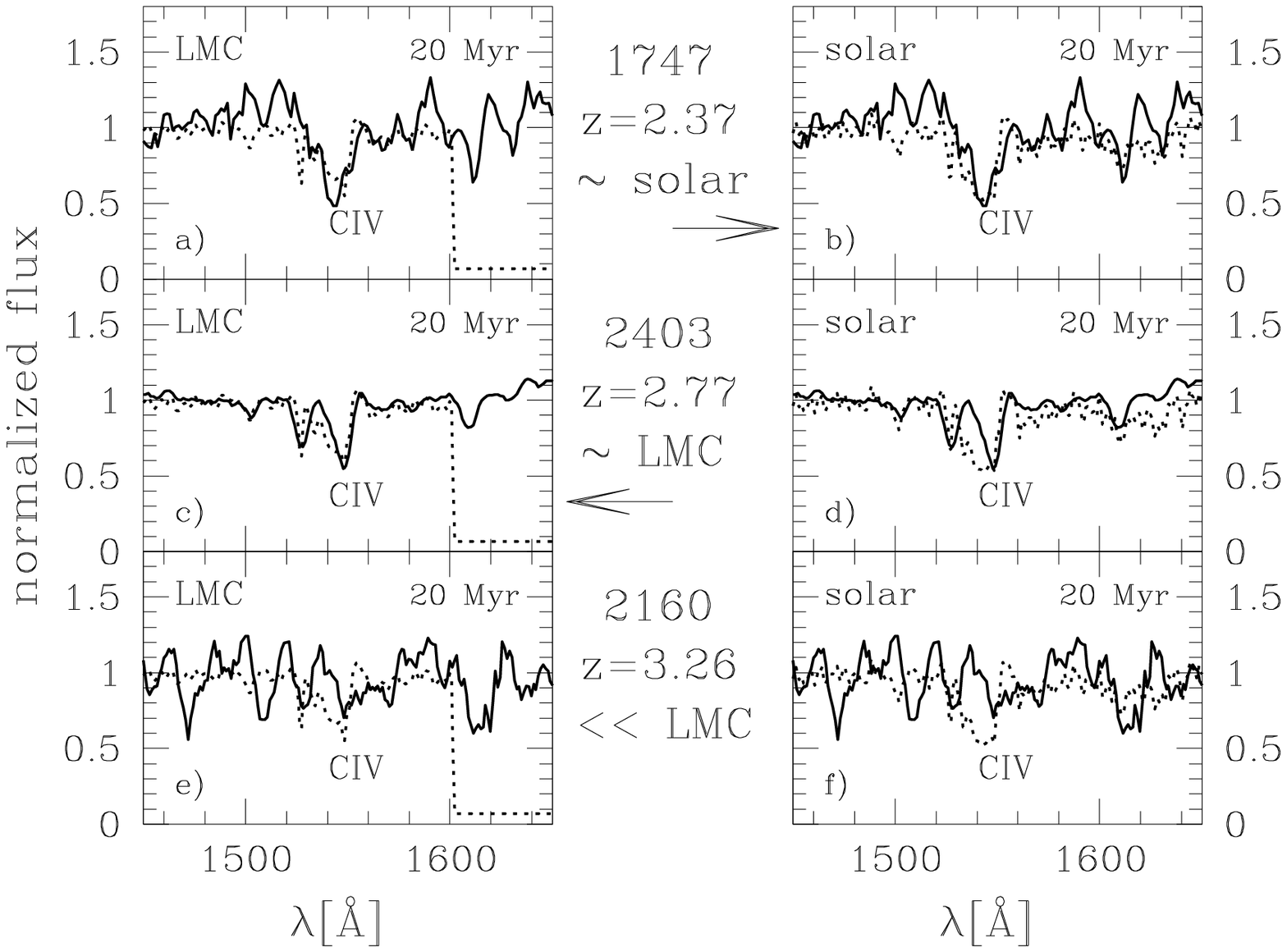}}
\end{center}
\vspace{-0.3cm}
{{\bf\scriptsize Fig. 2:}
\footnotesize
\em
Comparison between the observed spectra of 
galaxy 1747, 2403, 2160 (solid line) and synthetic spectra from 
Leitherer et al. (2001; dotted line). The synthetic spectra are based on
20 Myr old starbursts with  
continuous star formation (1\,M$_{\odot}/yr$) and the parameter 
$\alpha_{IMF} = 2.35$ and M$_{up}$ = 100\,$_{\odot}$. The left and 
right panels indicate models with LMC and solar metallicity, respectively.
}}

\noindent
Fig.~3a shows that for $z> 1$, $EW$(CIV) increases 
with decreasing redshift. A weighted $\chi^2$ fit gives a slope of 
$\alpha = -1.44 \pm 0.16$ (dashed line).  
Including the measured $EW$ of 5 nearby ($z=0$) starburst galaxies,
 arbitrarily chosen from the IUE archive, 
extends the trend to the local universe. A 
weighted $\chi^2$ fit gives a slope of $\alpha = -1.14 \pm 0.12$ 
(solid line).
\vbox{
\noindent
\begin{center}
{\includegraphics[width=5.8cm]{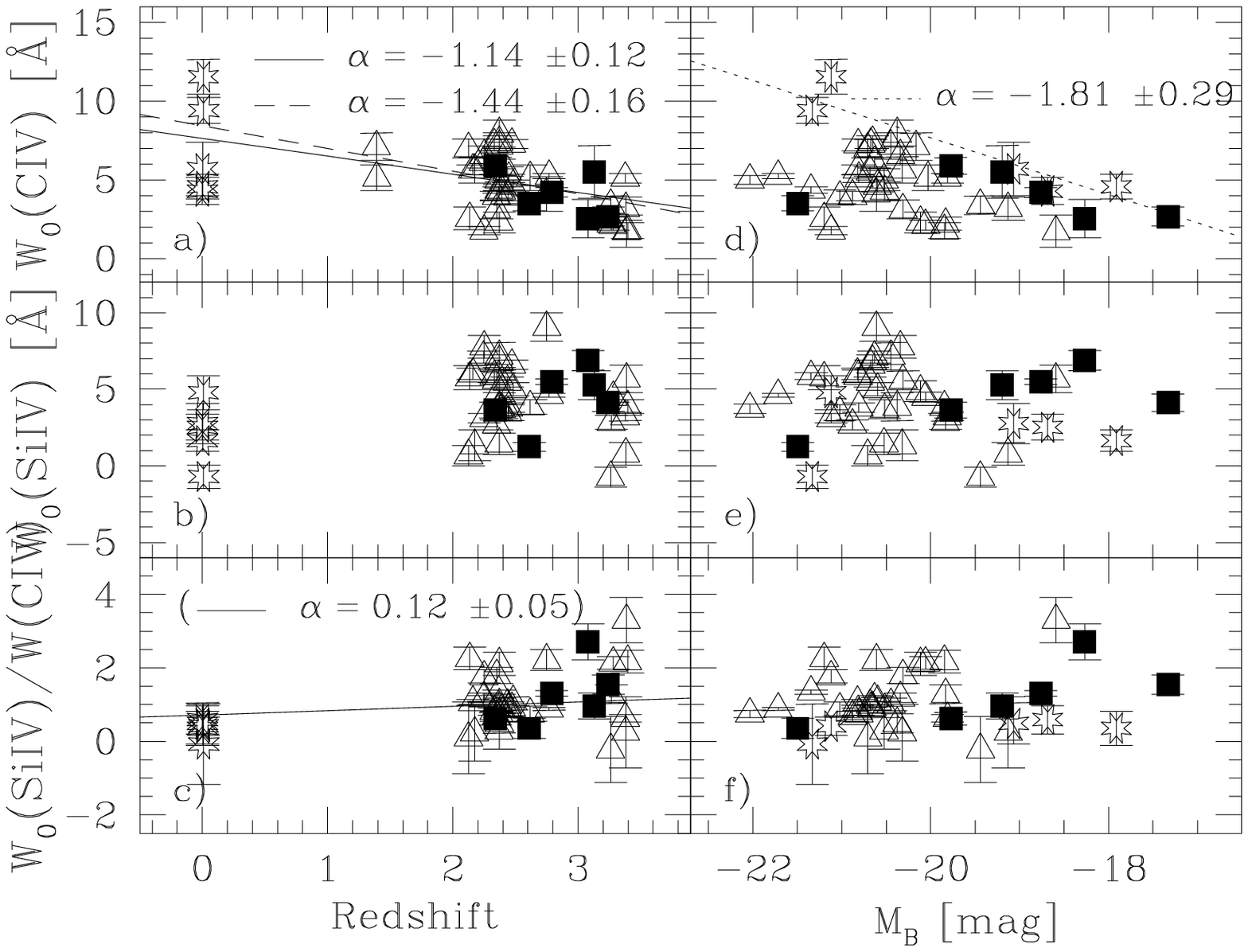}}
\end{center}
\vspace{-0.3cm}
{{\bf\scriptsize Fig. 3:}
\footnotesize
\em
Measured CIV~1550 and SiIV~1400 rest frame equivalent widths as 
well as their ratio SiIV/CIV\ versus redshift (a, b, c) and absolute 
B-magnitude (d, e, f). The absolute magnitudes are determined with 
$H_0 = 50\,\rm km/s/Mpc$ and $q_0=0.5$. For the $k$-correction 
we used values provided by M\"oller et al. (2001). 
Evolutionary corrections are not applied. 
Open triangles: FDF galaxies; filled squares:
Galaxies in the field of the cluster 1E0657, in the HDF-S and the AXAF 
Deep Field, respectively (see text);
stars: Nearby 
starburst galaxies. Dashed line, solid line, dotted line:
weighted $\chi^2$ for objects with $z>1$ only, for all shown galaxies 
and for the local ones ($z>1$), respectively.
}}

\noindent
Fig~4a and b show that both $EW$(CIV) and $EW$(SiIV)  
mainly depend on the metallicity of the stellar population but only 
little on the age of a starburst.
Hence the existing anticorrelation of $EW$(CIV) with 
$z$ indicates an increase of metallicity 
with decreasing redshift (i.e., increasing age of the universe). 
A similar behavior has been found for damped Ly${\alpha}$ systems 
(Savaglio et al. 2001).\\
\noindent
The fact 
that for SiIV\ no correlation with $z$ 
is present may be due to the fact that $EW$(CIV) 
is universally present for O~supergiants, main sequence stars and 
dwarfs, while $EW$(SiIV) is luminosity dependent and decreases rapidly from 
supergiants to dwarfs (Walborn \& Panek~1984; 
Pauldrach et al.~1990). 
\vbox{
\noindent
\begin{center}
{\includegraphics[width=5.8cm]{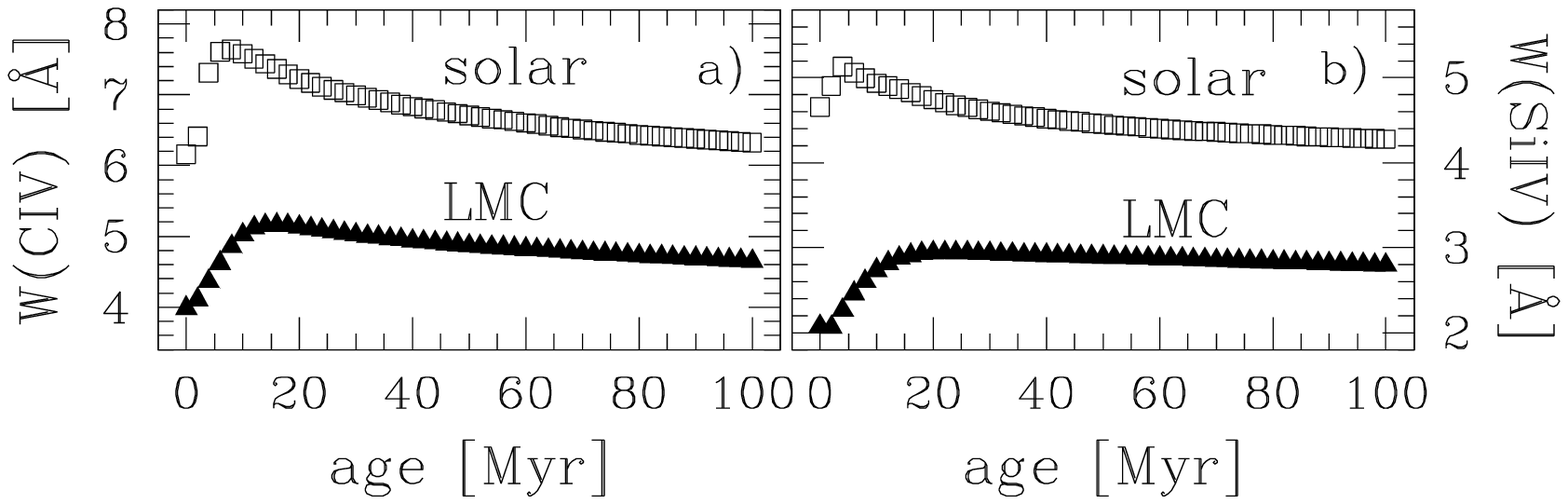}}
\end{center}
\vspace{-0.3cm}
{{\bf\scriptsize Fig. 4:}
\footnotesize
\em
Measured CIV~1550 (a) and SiIV~1400 (b) equivalent widths within 
the synthetic spectra of Leitherer et al. (2001) versus the age of the 
starburst. Open squares and filled triangles correspond to solar and LMC 
metallicity, respectively.
}}

\noindent
Therefore the SiIV\ is more strongly affected by population 
differences (i.e., stellar age differences) than the CIV\ line. 
$EW$(SiIV) has its maximum in B0/B1 stars, while $EW$(CIV) is mainly 
universally present in O and bright B~stars (e.g., Leitherer, et al. 1995). 
Hence the ratio of $EW$(SiIV)/$EW$(CIV) 
contains information about the star formation history (instantaneous 
or continuous) and the stellar population itself (e.g., IMF and cutoff 
masses).  Unfortunately the different parameters that determine the 
value of $EW$(SiIV)/$EW$(CIV) cannot be disentangled easily. 
As seen in Fig.~3c the ratios of $EW$(SiIV)/$EW$(CIV) show a weak 
trend corresponding to an increasing ratio with $z$ seems
 (a weighted $\chi^2$ fit gives $\alpha = 0.12 \pm 0.05$). 
This could be understood in terms of increase of the relative 
importance of continuous star formation (decrease of instantaneous 
starbursts) at low $z$.
\noindent
Finally, Figs.~3d to f show that there is no overall 
     dependence of the measured equivalent width on the galaxies' 
     luminosity. If, in fact, $EW$(CIV) is mainly determined by the 
     metallicity, the nearby starburst galaxies seem to follow the well 
     known local metallicity-luminosity relation (e.g. Kobulnicky \& 
     Zaritsky 1998), 
while the high-$z$ galaxies seem not to conform this 
     relation. As also found for Lyman break galaxies by Pettini et al. 
     (2001),  
the high-$z$ galaxies investigated in this paper seem to be 
     overluminous for their metallicity ($EW$(CIV)), which may indicate that 
     their mass-to-light ratios are low compared to present-day galaxies. 

\vspace{-0.4cm}
\section{Conclusions}
\vspace{-0.25cm}
$\bullet$ Our observed high-z galaxy spectra agree with synthetic ones.\\
$\bullet$ $EW$(CIV) is a good indicator for the galaxies metallicity.\\
$\bullet$ Our high-z starburst galaxies show increasing metal content 
with decreasing redshift and are 
overluminous for their metallicity compared with local starburst galaxies.\\

%

\end{multicols}
\end{document}